\documentclass[epj,epsf,psfig]{svjour}
\pdfoutput=1
\usepackage{array, graphicx, subfigure}
\usepackage{graphics}
\begin{document}
\title{A simple event weighting technique: }\subtitle {Optimizing the
measurement of the forward-backward asymmetry of Drell-Yan dilepton
and top-antitop pairs at hadron colliders} 
\author{A. Bodek
}                     
%
%
\institute{Department of Physics and Astronomy, University of
Rochester, Rochester, NY  14627-0171
}
\date{Received: date / Revised version: date}
%
\abstract{
We describe a simple technique  for optimizing the  extraction of
the forward-backward asymmetry ($A_{fb}$) of Drell-Yan lepton  pairs
($e^+e^-$,$~\mu^+\mu^-$) produced in $\bar{p}p$ and $pp$  collisions
at hadron colliders. The method employs simple event  weights which
are functions of the  rapidity  and  $|\cos\theta|$ decay angle of
the lepton pair. It  yields the best estimate of  the acceptance
corrected parton level ($\bar{q}q$)  forward backward asymmetry as a
function of final state dilepton mass ($M_{\ell\ell}$).   Typically,
when compared to the simple count method, the technique reduces the
statistical errors by  $20\%$ for $\bar{p}p$, and $40\%$  for   $pp$
collisions, respectively. The method is  equivalent to using the
maximum likelihood method, but is much easier to implement.
The  technique can be used to search   for new high mass and large
width Z' bosons which may be best detected   through the observation
of  deviations from  the Standard  Model expectation for the
forward-backward asymmetry.  The technique can also be applied in the
extraction of the foward-barckwad asymmetry in the production of
top-antitop pairs. 
} 
\maketitle
\section{Introduction}
\label{intro}

The Drell-Yan process  in which $q\bar{q}$ annihilations form
 intermediate $\gamma^{*}$ or $Z$ ($\gamma^{*}/Z$) vector bosons
decaying to
lepton ($e^+e^-$,$~\mu^+\mu^-$)  pairs  is particularly useful in
searching for new interactions at large
momentum transfers, $Q^{2}=M_{\ell\ell}^{2}$, where $M_{\ell\ell}$ is
the invariant mass of the lepton pair.
In leading order (LO) approximation, the momentum fractions $x_1$,
$x_2$ carried by the initial state quarks
and antiquarks in the proton and antiproton/proton, respectively, are
related to the rapidity $|y|$ \cite{cdf_coord}
 of the $\gamma^{*}/Z$ boson via the equation $x_{1,2} =
(M_{\ell\ell}/\sqrt{s})e^{\pm |y|}$, where $\sqrt{s}$ is 
 the center of mass energy. Dilepton pairs produced at large $|y|$
originate from collisions in which
one parton carries a large and the other a small momentum fraction
$x$.

Drell-Yan lepton  pairs  which are produced in  $q\bar{q}$
annihilations  display a 
forward-backward asymmetry because of the interference between photon
and $Z$ boson
exchange\cite{zp}.    This forward-backward asymmetry would be
modified by new resonances (e.g.
additional heavier $Z'$ bosons\cite{zp}) or new interactions at large
mass scales.

Although the mass limits  from  LEP~\cite{LEP} and Tevatron
($\bar{p}p$) experiments 
CDF~\cite{zprime} and
and \mbox{DO\hspace*{-1.5ex}/}~\cite{D0}
for a variety of $Z'$ models are in the  0.6-0.8~$TeV/c^2$ range, the
limits are
much lower  if the $Z'$ width (typically
$\Gamma_{Z'}\approx 0.01\cdot M_{Z'}$) 
is increased  to  account for the possibility of 
additional decays modes to
exotic fermions (which are predicted in $E_6$ models~\cite{E6}),
and/or 
supersymmetric particles. The limits are even lower if one includes
the possibility of a more general
model with enhanced couplings to the third generation.

Such a  $Z'$  (which has  larger width e.g.  $\Gamma_{Z'}=0.1\,
M_{Z'}$) would
produce only a small signal in  the dilepton mass spectrum because 
the total cross section is proportional to the square of the
amplitude. 
However, the change in the forward-backward asymmetry which  results
from
the interference with the standard model process is linearly
proportional to the
amplitude and would be observable as a change in the forward backward
asymmetry. This change will occur around the mass of the $Z'$  boson,
and also in some mass range
 below and above  the mass of the  $Z'$ boson\cite{widezp}.
\begin{figure}[ht]
\includegraphics[width=17pc,height=30pc]{{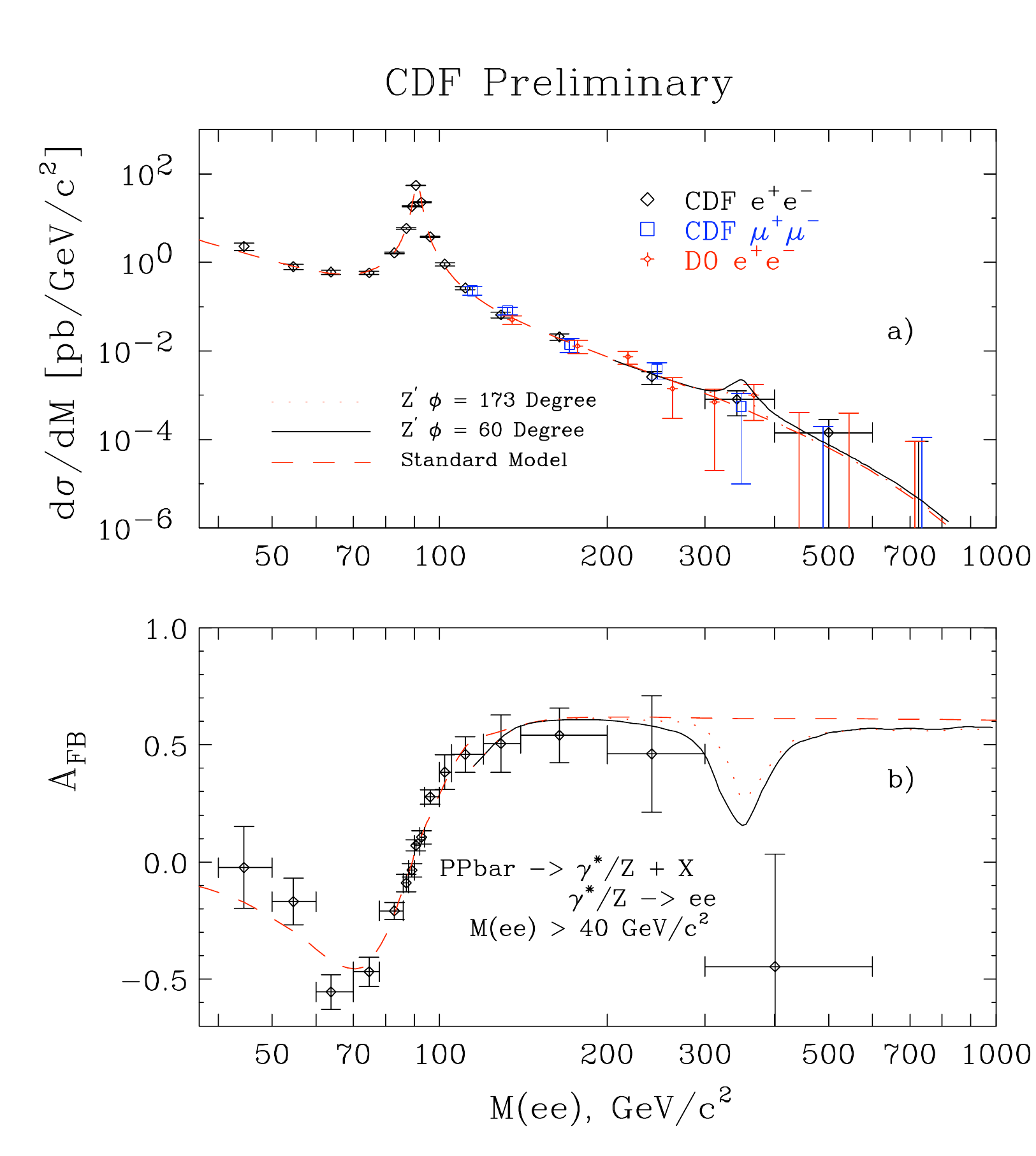}}
\caption{ Early Tevatron data (120 ~$pb^{-1}$). 
(a)  $d\sigma/dM_{\ell\ell}$ 
for   $e^+e^-$  pairs from CDF and DO,  
and $\mu^+\mu^-$ pairs from CDF. 
The Standard Model
theoretical predictions (dashed line) 
have been normalized (by a factor
 of 1.11) to the CDF data in the $Z$ boson mass region.
(b) CDF data for  $A_{FB}$ versus $M_{\ell\ell}$ ($e^+e^-$  pairs).
Also shown is the 
standard model expectation (dashed line) and the predictions 
for $d\sigma/dM$ and $A_{FB}$
with an extra $E_6$ boson  with $M_{Z'}=350~GeV/c^2$ and
$\Gamma_{Z'}=0.1 M_{Z'}$, for $\phi=60^0$ (solid) and 
$\phi=173^0$ (dotted).
}
\label{fig:1}
\end{figure}
%
This point is illustrated in  Figure 1 which is taken from
Reference\cite{widezp}.
The figure compares  early (120 ~$pb^{-1}$) high mass Drell-Yan 
$d\sigma/dM_{\ell\ell}$  data  from CDF and DO,  and  early $A_{FB}$
data from CDF 
to Standard Model theoretical predictions and  (as an example)  to a
prediction
with an extra $E_6$ boson  with $M_{Z'}=350~GeV/c^2$ and
$\Gamma_{Z'}=0.1 M_{Z'}$, for $\phi=60^0$ (solid) and 
$\phi=173^0$ (dotted).
As can be seen for this case, the signal in  $A_{fb}$ is larger than
the signal in 
$d\sigma/dM_{\ell\ell}$.
Since such new particles or new interactions may be best detected 
 through the observation of  deviations from  the Standard
 Model expectation for the forward-backward asymmetry, it is useful to
 devise experimental techniques to measure the forward-backward
asymmetry ($A_{fb}$)
 with the best possible precision.  Since fine mass bins ($\approx 25
GeV/c^2$)
 are required, the number of events per bin  at large $M_{\ell\ell}$
is small and the measurements are  statistically limited.
 
  In this communication we describe  a simple
  technique  for optimizing the  extraction of  the forward-backward
asymmetry
  in the production of Drell-Yan dilepton paris.  
The method employs simple event  weights,  which are functions of
the rapidity  and  $|\cos\theta|$ decay angle of the lepton pair. 
The method  yields the smallest statistical uncertainty in the
measurement 
of the  forward backward asymmetry as a function of $M_{\ell\ell}$.
It can be directly applied to  current  $\bar{p}p$ data at the
Fermilab Tevatron, as well as new
data that will be collected in pp
collisions at the Large Hadron Collider (LHC).   The method
 works well for both large and  small statistical samples.
 The method is  equivalent to using the maximum likelihood method,
but is much easier to implement.
 The technique can also be applied in the extraction of the
foward-backward asymmetry in the production of top-antitop pairs.

\section{ $q\bar{q}$ annihilations}

The differential cross-section for the parton level process for
$q\bar{q}$
annihilation can be written as

\begin{eqnarray}
\frac{d\sigma}{d( \cos\theta )} &=& A (1 + \cos^{2} \theta) + B'
\cos\theta \\
\end{eqnarray}

\noindent  where $\theta$ is the emission angle of the positive
lepton relative to the quark momentum in the center of mass frame,
and A and B' are parameters that depend on the weak isospin and
charge of the incoming fermions.  (We use  B'  for 
coefficient of the  $cos\theta$ at the parton level, and B for the
coefficient of the  $cos\theta$  at the nucleon level). 

The cross sections for forward ($for-$) events ($\sigma_{F}$) and
backward ($back-$)  events ($\sigma_{B}$) are given by

\begin{eqnarray}
\sigma_{f} &= &\int_{0}^{1} \frac{d\sigma}{d(\cos\theta)}
d(\cos\theta) \\\nonumber
& =& A\left(1+\frac{1}{3}\right) +  B'\left(\frac{1}{2}\right) \\
\sigma_{b}&=& \int_{-1}^{0} \frac{d\sigma}{d(\cos\theta)}
d(\cos\theta) \\\nonumber
& =&  A\left(1+\frac{1}{3}\right) -  B'\left(\frac{1}{2}\right)
\nonumber
\end{eqnarray}

The electroweak interaction introduces the asymmetry (a linear
dependence on $\cos\theta$), which can be expressed as

\begin{eqnarray}
A_{fb} ^{total}& =&  \frac{\sigma_{f}-\sigma_{b}}
{\sigma_{b}+\sigma_{b}} = \frac{3B'}{8A}
\end{eqnarray}

For $\bar{p}p$ collisions  (e.g. at the Tevatron), the direction of
the quark is predominately in the proton direction, and  the
direction of the antiquark is predominately in the antiproton
direction.  Therefore, the
forward backward asymmetry for  $q\bar{q}$ processes is  easy to
measure in $\bar{p}p$ collisions. 

If  $N_{f}$ is in number of events in the forward direction of the
quark and $N_{b}$ is the number of events in the backward direction
of the quark we obtain the following expression for the total forward
backward-asymmetry ($A_{fb}^{total}$) and its error  ($\Delta
A_{fb}^{total}$): 
\begin{eqnarray}
\left[ A_{fb} \right]^{total}&=& \frac{N_{f}-N_{b} }{N_{f}+N_{b}} =
\frac{N_{f}-N_{b} }{N}\\ \nonumber
{\frac {N_{f} } {N_{b} } } &=&  \frac {1- A_{fb}^{total} } {1+
A_{fb}^{total} }\\ \nonumber
N_{f} & = &  \frac{1+ A_{fb} ^{total}}{2} N \\ \nonumber
N_{b}& = &  \frac{1 - A_{fb} ^{total}}{2} N \\\nonumber
\label{error1}
 \Delta A_{fb}^{total}  &=&  \frac{2}{N} \left[  \frac{N_{f}N_{b}}{N}
\right]^{1/2}\\
 \label{error2}
 \Delta A_{fb}^{total} &=& \left[
\frac{1-(A_{fb(expected)}^{total})^{2}} {N} \right]^{1/2}
\end{eqnarray}

where we have used $\Delta N_{f}= (N_{f})^{1/2}$ and  $\Delta N_{b}=
(N_{b})^{1/2}$,
and $N=N_f+N_b$.  Since for
Poisson statistics\cite{bodekNIM}, the fractional error is
$(1/N_{expected})^{1/2}$ and not  $(1/N_{observed})^{1/2}$ , we 
use $A_{fb(expected)}$ in equation \ref{error2}.  For $\bar{p}p$
collisions above the $Z$ mass peak,  $A_{fb(expected)}$=0.6.
 In this region,   $\Delta A_{fb} = 0.800 \cdot (1/N)^{1/2}$.  
 
 Therefore,  a  measurement  with 100 events yields
 a statistical  error of 0.08.  
 This level of precision is needed to observe the deviation from the
Standard Model for the $Z'$  example shown in figure 1.  Later in
this paper we show that a reduction in the error  (of about $20\%$)
can be obtained by using the information in the angular distribution
of the forward and backward events.

\section{Quarks bound in a nucleon}
 
When quarks are bound in the nucleon, the dilepton can be produced
with non-zero transverse momentum.  This is described in 
detail in appendix A.  For $p\bar{p}$ or $pp$ collisions the angular
distribution
of $\gamma^{*}/Z$ vector bosons decaying to
$e^+e^-$ or $~\mu^+\mu^-$ pairs  is given by:
\begin{eqnarray}
\frac{d\sigma}{d( \cos\theta )} &=& A [1 + \cos^{2} \theta
+h(\theta)]+ B \cos\theta \\
\label{qterm}
h(\theta)  &=& \frac {1}{2}  A_{0}(M_{\ell\ell}, P_{T}) (1 - 3
\cos^{2} \theta )
\end{eqnarray}
 The $q\bar{q}$  center of mass frame is well defined when the lepton
pair has zero transverse momentum ($P_{T})$. For a non-zero
transverse momentum of the dilepton pair, the $q\bar{q}$  center of
mass frame is approximated by the Collins-Soper frame\cite{cs}.

The term  $h(\theta,M_{\ell\ell}, P_{T})$ is a small QCD correction
term  which is zero when the transverse momentum of the dilepton pair
is zero. As shown in Appendix A, the angular coefficient $A_0$ is
only a  function of the dilepton mass ($M_{\ell\ell}$) and
transverse momentum ($P_{T}$).   The $h(\theta,M_{\ell\ell},P_{T})$
term integrates to zero when the cross section is integrated over all
$\cos^{2} \theta$.  
%
and increases with transverse momentum. 
For simplicity  we will use $A_{0}=0$ (e.g.  $P_{T}$=0) in our
calculations.
  In order to include the dependence on the $h(\theta,M_{\ell\ell},
P_{T})$ term, we  can use the 
  following function derived in Appendix A:
\begin{eqnarray}
 A_0=  \frac {P_T^2/M^2_{\ell\ell}} {1+ P_T^2/M^2_{\ell\ell}}
\label{Afit}
\end{eqnarray}
 
QCD corrections also dilute the forward-backward asymmetry as follows:
 \begin{eqnarray}
 \frac {B}{B'} &=&   \ \frac {1} {(1+ P_T^2/M^2_{\ell\ell})^{1/2}}
 \label{Bfit}
\end{eqnarray}
 On average,  both QCD correction terms are very small. 
  For Drell-Yan events  the average ratio $P_T/M_{\ell\ell}$ is about
0.13 (e.g. 12 GeV $P_T$
 at the $Z$ mass of 90 GeV). This implies that on average
$<A_0(P_T)>= 0.018$, and   $<\frac {B}{B'}(P_T>=0.991$. 
  
The emission of photons by the final state leptons also distorts the
angular distribution (in addition to smearing events to lower mass).
The  effects of final state QED photons  are best  accounted for as
part of an overall radiative correction.
Alternatively, the angular distortion from QED radiation  can also be
corrected  for by using  effective  angular coefficients which
combine the QCD and QED corrections  (expressions for the  larger
$A_0^{rad}(P_T,M_{\ell\ell})$ and larger $\frac
{B^{rad}}{B'}(P_T,M_{\ell\ell})$ are given in Appendix A).

\section{ Misidentification of the quark direction in $pp$ collisions}

For $\bar{p}p$ collisions the direction of the quark is primarily
along the direction of the proton. However,  there is a  a small
probability for a misidentification ($misID$)  that originates from
the charge misidentification  probability in the tracker. There is
also an  additional small  $misID$ that comes from the  small
fraction of events in which a sea antiquark  in the proton interacts
with a sea quark in the antiproton.  This $misID$ is very small for
large dilepton final state mass. 

We now turn to the measurement of the forward-backward asymmetry in
$pp$ collisions (e.g. at the Large Hadron Collider). 
Although $pp$ collisions are symmetric, there is still a forward
backward asymmetry
if the quark direction is defined to be the direction of motion of
the Drell-Yan pair.  This originates from
the fact that  on average, quarks carry a larger fraction of the
proton's momentum than antiquarks. 
However, there is a significant $misID$ that originates from  the
fraction of events for which the antiquark carries a larger fraction
of the momentum than the quark. This $misID$  dilutes the observed
asymmetry.

In most theoretical studies of the production of new $Z'$ bosons in
$pp$ collisions at the LHC, 
this dilution is included in  calculation for the prediction for the
observed forward-backward asymmetry.
Here,  we show that we can obtain a higher sensitivity to new
particle searches by 
 correcting  the data for the $misID$ fraction  on an event by event
basis. 
 
At  small rapidity the misidentification probability $w_{i}$  is
large ($0.5$ at $|y|=0$).  At large rapidity the misidentification
probability $w_{i}$  is small.  We show below  that  by taking this
information into account   we can reduce the error on the extracted
$q\bar{q}$ asymmetry.  

\section{ Correcting for $misID$ in $pp$ collisions}

We illustrate this point for the case of a  high statistics
measurement of the forward backward asymmetry in a specific
mass bin (e.g $450~GeV/c^2$).   

We first extract the $q\bar{q}$ asymmetry by correcting the measured
asymmetry for the average $misID$ probability (this commonly used
method is called the event count method).

   We then show that we reduce  the error on the  extracted
quark-antiquark  asymmetry by binning the data in ten rapidity bins
and fitting for the weighted average of the extracted parton level
asymmetries from all of the ten  rapidity bins. 
   
    Then we derive an event weighting technique that is equivalent to
the fit method in the high statistics case, but which can also be
used in the limit of very small statistical samples.

For proton-antiproton collisions we define the $misID$  probability
($w_{i}$)  for each  $y_{i}$ bin  
as the fraction of events for which the antiquark carries a larger
fraction of the momentum
than the quark.  (For proton-antiproton collisions $w_{i}$ is the
fraction of events
for which the interactions occurs between a sea quark in the
antiproton
and a sea antiquark in the proton).

\begin{equation}
x_{1(2)} = \frac{M_{\ell\ell}}{\sqrt{s}}\times{e^{\pm |y_i|}}
\label{w_pdf}
\end{equation}
\begin{equation}
w_{i} \approx \frac{ \sum_{flavor} { v_{q} \{
q(x_{2})\cdot{\bar{q}(x_{1})}} \} } { \sum_{flavor}{ v_{q}
\{q(x_{2})\cdot{\bar{q}(x_{1})}+q(x_{1})\cdot{\bar{q}(x_{2})}} \} }
\end{equation}

Where   $M_{\ell\ell}$  is the dilepton mass  ($M_{\mu\mu}$ or
$M_{ee}$),  and  $|y_i|$ is the rapidity of the
dilepton pair.  Here $q(x)$ denotes the quark distributions  ($u(x)$,
$d(x)$ , $s(x)$ , $c(x)$,  $b(x)$)  and  $\bar{q}(x)$ denotes the
antiquark distributions ($\bar{u}(x)$,  $\bar{d}(x)$,
  $\bar{s}(x)$,   $\bar{c}(x)$,  $\bar{b}(x)$) for the 
various flavors in the nucleon.  The parameter $v_{q}$ denotes the
$Z/\gamma$  couplings of to each flavor (which  a function of the
dilepton mass).  At large $M_{\ell\ell}$ the   $u(x)$ and
${\bar{u}(x)}$ quark distributions dominate the expression for the
dilution factor.

Note that the $misID$ in the forward-backward asymmetry should
include the interference between photon and $Z$ boson exchange for
each quark flavor, which in general is a more complicated function of
the couplings.  In addition, the  $misID$ is affected by radiative
emission of photons and detector resolution.   Therefore, it is best
to use a Monte Carlo generator (such as {\small  PYTHIA}\cite{pythia}
or {\small ZGRAD2}\cite{zgrad}) to empirically determine the
functional dependence of
the dilution factor $w_i$ as a function of the two measured
variables  $y_{\ell\ell,measured}$ and  $M_{\ell\ell,measured}$. 

\begin{equation}
w_{i} = f (y_{\ell\ell,measured},M_{\ell\ell,measured}). 
\end{equation}

 We now proceed to  correct for the $misID$ and extract the
$q\bar{q}$  forward-backward asymmetry
 for each one of the ten $y_{i}$ rapidity bins.  In the expressions
below,   $n_{f,i}$ and $n_{b,i}$ are the defined as the measured
(i.e.  diluted) number of forward events and backward events  in each
bin, 
and   $N_{f,i}$ and $N_{b,i}$  are defined as the number of true
forward and true backward events (for  $q\bar{q}$ collisions) in the
bin.  

For a given $y_{i}$ rapidity bin with a  misidentification
probability $w_{i}$ 
the measured and true number of forward and backward events are
related by the
following expressions .

\begin{eqnarray}
n_{f,i} &=&  N_{f,i} (1-w_{i})  + N_{b,i} (w_{i})\\ \nonumber
n_{b,i} &=&  N_{b,i} (1-w_{i}) + N_{f,i} (w_{i})\\  \nonumber
N_{f,i} &=&  n_{f,i} (1-w_{i})/L_{i}  - n_{b,i} (w_{i})/L_{i}\\
\nonumber
N_{b,i} &=&  n_{b,i} (1-w_{i})/L_{i}  -n_{f,i} (w_{i})/L_{i}
\nonumber
\end{eqnarray}

where $L_{i}$ = $ (1-2w_{i})$ is defined as the dilution factor.  For
a $|y_i|$ rapidity bin, the corrected parton level asymmetry is given
by
\begin{eqnarray}
A_{fb-i}^{total} &=& \frac{N_{f,i}-N_{b,i} }{N_{f,i}+N_{b,i}}  \\
\nonumber
N_{f,i}-N_{b,i} &=&n_{f,i}/L_{i} -n_{b,i}/L_{i}\\ \nonumber
N_{f,i}+N_{b,i}&=&n_{f,i}+n_{b,i}=n_i \nonumber
\end{eqnarray}

which yields 
\begin{eqnarray}
\label{dilution}
A_{fb,i} ^{total}&=& \frac{1}{L_{i}} \frac{n_{f,i}-n_{b,i} }{n_{i}}
\\ 
\left[ \Delta A_{fb-i}^{total}\right] &=&  \frac{1}{L_{i}}
\frac{2}{n_i} \left[  \frac{n_{f,i}n_{b,i}}{n_{i}}  \right]^{1/2} 
\end{eqnarray}
where we have used $\Delta n_{f,i}= (n_{f,i})^{1/2}$ and  $\Delta
n_{b,i}= (n_{b,i})^{1/2}$.

We find that  a measurement of the $q\bar{q}$ asymmetry in the case
where there is   $misID$ probability of $w_{i}$ results in an
increase of the error in the extracted parton level asymmetry  by a
factor of $1/L_{i}$  (which is equivalent to reducing the number of
events by a factor of $L_{i}^{2} = (1-2w_{i})^2$).

  If we want to combine different $|y_{i}|$ bins together, we need to
weight the events by the inverse of the square of the statistical
error
 in each bin.  This is achieved by multiplying the expressions for
$N_{f,i}$ and $N_{b,i}$ by   $L_{i}^{2}$.  Since this factor appears
both in the numerator and denominator of the expression for
$A_{fb-i} ^{total}$, it does not change the extracted value or error
of the parton level asymmetry.
   However, when we combine $|y_{i}|$ bins together using event
weighting,  
 this factor accounts for the difference in statistical errors
between the $|y_i|$  bins as follows.

 \begin{eqnarray}
    k_{1,i} &=& (1-w_{i})(1-2w_{i})\\   \nonumber
k_{2,i} & =&  (w_{i})(1-2w_{i}) \\  \nonumber
    N_{total} & =&  \sum_{all-events}\left[  1 \right] \\  \nonumber
S_{f} &=&  \sum_{for-events}  k_{1,i}  -  \sum_{back-events} k_{2,i}
\\  \nonumber
\left[ \Delta S_{f} \right]^{2} &=&  \sum_{for-events}   k^{2}_{1,i}
+ \sum_{back-events} k^{2}_{2,i}   \\  \nonumber
S_{b} &=& \sum_{back-events}    k_{1,i}  -  \sum_{for-events} k_{2,i}
\\  \nonumber
\left[ \Delta S_{b} \right]^{2} &=& \sum_{back-events}   k^{2}_{1,i}
+\sum_{for-events}  k^{2}_{2,i}  \\  \nonumber
A_{fb}^{total} &=& \frac{S_{f}-S_{b} }{S_{f}+S{b}} \nonumber
\end{eqnarray}
 
Now  
$\left[ \Delta S_{f} \right]$ and $ \left[ \Delta S_{b} \right] $ are
correlated with each
other in a complicated way. In order to simplify the calculation of
the error, we  combine terms to isolate
sums which are for forward events,  and sums which are for backward
events, separately as follows:

\begin{eqnarray}
    k_{A,i} &=&    k_{1,i} -   k_{2,i} =(1-2w_{i})^{2}\\  \nonumber
k_{B,i} & =&    k_{1,i}  +   k_{2,i} =  (1-2w_{i}) \\ \nonumber
   N_{total} & =&  \sum_{all-events}\left[  1 \right] \\ \nonumber
A &=& S_{f} +S_{b} =  A_{1}+A_{2}\\  \nonumber
B &=& S_{f} - S_{b} =  B_{1}-B_{2}\\  \nonumber
A1 &=&  \sum_{forward-events}  k_{A,i}   \\  \nonumber
A2 &=&  \sum_{back-events} k_{A,i}   \\  \nonumber
B1 &=&  \sum_{forward-events}  k_{B,i}    \\  \nonumber
B2 &=  & \sum_{back-events} k_{B,i}   \\  \nonumber
\left[ \Delta A_{1}\right]^{2} &=&  \sum_{forward-events}  k_{A,i}
^{2}  \\ \nonumber
\left[ \Delta A_{2} \right]^{2} &=&  \sum_{back-events} k_{A,i} ^{2}
\\ \nonumber
\left[ \Delta B_{1}\right]^{2} &=&  \sum_{forward-events}
k_{B,i}^{2}    \\ \nonumber
\left[ \Delta B_{2} \right]^{2} &=  & \sum_{back-events} k_{B,i} ^{2}
\\ \nonumber
A_{fb}^{total}  &=&    \frac{B} {A}  =   \frac{B_{1}  - B_{2} }
{A_{1} + A_{2}} \nonumber
\end{eqnarray}

Now  $\Delta A_{1}$ is $100\%$ correlated with $\Delta B_{1}$   and
$\Delta A_{2}$ is $100\%$ correlated with $\Delta B_{2}$ . 
We handle these correlations  as follows.  
\begin{eqnarray}
\Delta A_{1} & = &  \Delta B_{1} \cdot \frac{A1}{B1} \\ \nonumber
\Delta A_{2} & = &  \Delta B_{2} \cdot \frac{A2}{B2 } \\ \nonumber
 \left[ \Delta A_{fb} ^{total}\right] ^{2}  &=&  
\frac {1}  {(A_{1}+A_{2})^{4}}  \left[ E_{1}^{2} + E_{2}^{2}  \right]
\\ \nonumber
E_{1}^{2} & = &    \frac  {\left[ \Delta B_{1}
\right]^{2}}{B^{2}_{1}} (A2  B1 +A1  B2 )^{2}\\ \nonumber
E_{2}^{2}&=&   \frac {\left[ \Delta B_{2} \right]^{2}}{B^{2}_{2}}
(A2  B1 +A1  B2 )^{2} \nonumber
\end{eqnarray}

\begin{table*}[htb]
\caption {Proton-Proton collisions: Numerical example of measuring
$A_{fb}$  in  ten bins of  $|y|$ for
the range of $|y|$ between 0  and 2.0.
For this study we assume an asymmetry $A_{fb}$ = 0.6 which is
measured with 1000 events (for $0<|y|<2$)
in $pp$ collisions at the LHC.
We assume that the differential cross section is constant in $|y|$.
The $misID$ probability
is assumed to be  $w(|y|) = 0.5 (2-|y|)/2$ which is a simple
approximation to  to the
$misID$ values  for a dilepton mass of 450 $GeV/c^2$.
 In each range in $|y|$ we compare the error in the extracted
$\bar{q}q$  asymmetry from  a simple count
 of events, the error using a least square fit
to the values extracted for all  $|y_i|$ bins, and the error from the
proposed event weighting formula. The
 error using a simple count is about $20\%$ larger than the
error using  the the proposed event weighting scheme. The last
column shows the error for the case
of $\bar{p}p$ collisions (with a $misID$=0).  }
\newcommand{\m}{\hphantom{$-$}}
\newcommand{\cc}[1]{\multicolumn{1}{c}{#1}}
\renewcommand{\tabcolsep}{1pc} 
    \begin{center}
\begin{tabular}{|c||c||c||c||c||c|c||}
\hline
{$|y| -range$}  	& {$N$} 	&{$simple~Count$} 	& {$|y|~bins~fit$}	&
{$Weights$}
 &  {$Improv.$} & {$\bar{p}p$}\\
{$range$}  &{$bins$} & {$Error~pp$} &  {$Error~pp$}& {$Error~pp$} &
{$Factor~pp$}& {$Error$} \\
\hline\hline
0.-0.2&	1	&	1.999	&	1.9991	& 1.9991	        &	1.00 & 0.087	\\
0-0.4	&	2	&	0.706	&	0.6301	&	0.6298	&	1.12  & 0.061	\\
0-0.6	&	3	&	0.383	&	0.3350	&	0.3346	&	1.14 & 0.050	\\
0-0.8	&	4	&	0.248	&	0.2145	&	0.2141	&	1.15  & 0.043	\\
0-1.0	&	5	&	0.177	&	0.1512	&	0.1511	&	1.16	 & 0.039\\
0-1.2&	6	&	0.134	&	0.1136	&	0.1131	&	1.17	 & 0.035\\
0-1.4&	7	&	0.106	&	0.0886        &	0.0882       &	1.18  & 0.033	\\
0-1.6&	8	&	0.086	&	0.0711	&	0.0706	&	1.21	 & 0.031\\
0-1.8&	9	&	0.071	&	0.0582       &	0.0578	&	1.23 & 0.029	\\
0-2.0	&	10	&	0.060	&	0.0483        &	0.0479	&	1.26	 & 0.027 \\
\hline\hline
\end{tabular}
\label{table1}
    \end{center}
\end{table*}

A specific numerical example is shown in Table \ref{table1}.   Here
we show
the case of a  measurement of   $A_{fb}$  in  ten bins of  $|y|$ for
the range of $|y|$ between 0  and 2.0.
For this study we assume that an asymmetry of  $A_{fb}$ = 0.6  is
measured with 1000 events (for $0<|y|<2$)
in $pp$ collisions at the LHC.
We assume that the differential cross section is constant in $|y|$.
The $misID$ probability
is assumed to be  $w(|y|) = 0.5 (2-|y|)/2$ which is a simple
approximation to  to the
$misID$ values  for a dilepton mass of 450 $GeV/c^2$ at the LHC.
 In each  range in  $|y|$ we compare the error in the extracted
$\bar{p}p$ asymmetry from  a simple count
 of events;  the error extracted from a least square fit
to the values extracted from each $|y_i|$ bin, and the error from the
our event weighting formula. 
As expected, the error from the least square fit to the ten $|y_i|$
bins is the same as
the error from our weighting formula. 
The error using a simple count is about $20\%$ larger than the
error using  the event  weighting scheme. The last
column shows the error for the case of $\bar{p}p$ collisions (with a
$misID$=0).

Our method is equivalent to combining different $|y_{i}|$ bins
together by weighting the events by the inverse of the square of the
"expected"  statistical  error in each bin.  This has been done  by
multiplying the expressions for $N_{f,i}$ and $N_{b,i}$ by
$L_{i}^{2}$.
Weighting by a  statistical error which is based on the  "expected"
number of events in a bin
(instead of the measured number of events in a bin)  is more correct,
and is equivalent\cite{bodekNIM} to using
the maximum likelihood technique for combining bins. Therefore, our
event weighting technique is
equivalent to using a maximum likelihood technique, but is much
easier to implement.

\begin{table*}[htb]
\caption {Proton-Antiproton collisions: Numerical example for an
asymmetry $A_{fb}$ = 0.6 measured with 1000 events.  Here
$A_{fb}^{total}$ is measured in  ten bins of  $|\cos\theta|$.
 In each $|\cos\theta_j|$ bin we compare the error from the standard
error formula, and the error from the event weighting formula. As
expected, the two yield identical results.
 In addition, we show a comparison of  the average $A_{fb}^{total}$
 of all ten  $|\cos\theta_j|$ bins  calculated  two different ways.
The error in the average extracted from  a least square fit to the 10
$A_{fb-j}^{total}$ values (0.0196)  is close to the error in the
average  determined from the weighted sum of all the events (0.0210).
The error in $A_{fb}^{total}$  from the weighted sum of all the
events  is $20\%$ lower than the error of 0.0253 obtained  from a
simple count of all forward and backward events. }
 \label{table2}
\newcommand{\m}{\hphantom{$-$}}
\newcommand{\cc}[1]{\multicolumn{1}{c}{#1}}
\renewcommand{\tabcolsep}{1pc} 
   \begin{center}
\begin{tabular}{|c||c||c||c||c||c||c||}
\hline
{$|c_{j}|=$} & {$n_{f}$} &  {$n_{b}$} & {$A_{fb-j}(|c_{j}|)$} &
{$A_{fb-j}^{total}$} & {$\Delta A_{fb-j}^{ total}$}  &{$\Delta
A_{fb-j}^{ total}$} \\
{$|\cos\theta_{j}|$}& &   & {$measured$} &{$extracted$} &
{$|\cos\theta_{j}|~bin$}  &{$from~weights$} \\
\hline\hline
0.05	&	41	&	34	&	0.080	&	0.60	&	0.864	&	0.864	\\
0.15	&	47	&	29	&	0.235	&	0.60	&	0.284	&	0.284	\\
0.25	&	55	&	25	&	0.376	&	0.60	&	0.165	&	0.165	\\
0.35	&	63	&	21	&	0.499	&	0.60	&	0.114	&	0.114	\\
0.45	&	72	&	18	&	0.599	&	0.60	&	0.085	&	0.085	\\
0.55	&	82	&	16	&	0.676	&	0.60	&	0.066	&	0.066	\\
0.65	&	92	&	14	&	0.731	&	0.60	&	0.054	&	0.054	\\
0.75	&	104	&	14	&	0.768	&	0.60	&	0.045	&	0.046	\\
0.85	&	116	&	14	&	0.790	&	0.60	&	0.041	&	0.041	\\
0.95	&	128	&	14	&	0.799	&	0.60	&	0.038	&	0.028	\\
\hline
all	&	800	&	200	&	0.6	&	0.60	&	0.0196	&	0.0210	\\
\hline
all	&	800	&	200	&	0.6	&	0.60	&	simple-count	&	0.0253	\\
\hline
\end{tabular}
   \end{center}
\end{table*}
 
  \begin{table*}[htb]
\caption {Proton-Antiproton collisions:  Numerical example of
measuring  $A_{fb}$  for
different acceptance ranges in  $|\cos\theta|$. 
The example is for an asymmetry $A_{fb}$ = 0.6 measured with a  total
of 1000 events
(for all values of   $|\cos\theta|$). 
 For each $|\cos\theta|$ range we show the error in $A_{fb}$ from a
simple count,  the error from the event weighting procedure, and  the
error from fitting
bins in $|\cos\theta||$ (which is very similar to event weighting).
Also shown is the improvement factor  in the error when the event
weighting procedure 
is used (versus a simple count).  For a typical range of
$|\cos\theta|$, using the simple weighting formula leads to  more
than $20\%$ reduction in the error.  }
\label{table3}
\newcommand{\m}{\hphantom{$-$}}
\newcommand{\cc}[1]{\multicolumn{1}{c}{#1}}
\renewcommand{\tabcolsep}{1.1pc} 
   \begin{center}
\begin{tabular}{|c||c||c||c||c||c||cl|}
\hline
{$|\cos\theta|$}  	& {$N$} 	&{$simple~Count$} 	&
{$|\cos\theta|~bins$}	& {$Event~Weights$}
 &  {$Improvement$}\\
{$range$}  &{$bins$} & {$Error$} &  {$fit~Error$}& {$Error$} &
{$factor$} \\
\hline
0-0.1 	&1	& 0.8642	&	0.8642	&	0.8642	 & 1.000	\\
0-0.2		&2	&0.3042	&	0.2796	&	0.2687	 & 1.132	\\
0-0.3		&3	&0.1644	&	0.1410	&	0.1403	 & 1.172	\\
0-0.4		&4	&0.1058	&	0.0884	&	0.0881 	&1.201	\\
0-0.5 	&5	&0.0749	&	0.0611	&	0.0611 	&1.225	\\
0-0.6		&6	&0.0563	&	0.0459	&	0.0454	 & 1.241	\\
0-0.7		&7	&0.0442	&	0.0346	&	0.0354	& 1.247	 \\
0-0.8		&8	&0.0358	&	0.0277	&	0.0288 	& 1.242	\\
0-0.9		&9	&0.0298	&	0.0230	&	0.0242	& 1.227	 \\
0-1.0		&10	&0.0253	&	0.0196 	&	0.0210	& 1.205	\\
\hline \hline
\end{tabular}
   \end{center}
\end{table*}

 \section{Including information  in the angular distribution in
$\bar{p}p$ collisions}

We now investigate how much can be gained by looking at the asymmetry
in bins of  $|c_{j}|= |\cos\theta_{j}|$. We start with the case of
$\bar{q}q$ collisions and divide the sample into ten bins in
$|\cos\theta_{j}|$.
The asymmetry as a function of $|c_{j}|$ bin  is:

\begin{eqnarray}
A_{fb-j} (c_{j}) & =& \frac{\sigma_{f}(|c_{j}|)-\sigma_{b}(|c_{j}|)}
{\sigma_{f}(|c_{j})|+\sigma_{b}|(c_{j}|)}\\ \nonumber
A_{fb-j} (c_{j}) & =& \frac {N_{f,j}-N_{b,j} }   {N_{f,j}-N_{b,j} }
\\ \nonumber
&= &\frac{B|c_{j}|}{A(1+c_{j}^{2}+h(M_{\ell\ell},\theta, P_{T})}\\
\nonumber
&  = & A_{fb-j}^{total} \left[ \frac{8|c_{j}|}
{3(1+c_{j}^{2}+h(\theta)} \right] \nonumber
\end{eqnarray}

At $|c_{j}|=|\cos\theta_{j}|=0$, the measured asymmetry $ A_{fb-j}
(0)$=0.  At  $|c_{j}|$ = $|\cos\theta_{j}|$ = $0.45$,  the measured
asymmetry $A_{fb-j} (0.45)=A_{FB}^{total}$.
At  $|c_{j}|=|\cos\theta_{j}|=1$,   the asymmetry    $A_{fb}(1)$ =
(4/3)$A_{fb}^{total}$.  The measured asymmetry in  each $|c_{j}|=
\cos\theta_{j}$ bin  can be related to the total (integrated over all
$\cos\theta$) asymmetry. Therefore
each $|\cos\theta_{j}|$  bin provides an independent measurement
$A_{fb-j}^{total}$ of the  total asymmetry.

\begin{eqnarray}
A_{fb-j}^{total} &= & \frac {3}{8}  \cdot \frac{N_{f,j}-N_{b,j}
}{N_{f,j}+N_{b,j}} \cdot \frac{1}{M_{j}} \\ \nonumber
\Delta A_{fb-j}^{ total} &=& \frac {3}{8M_{j}}
\frac{2}{N_{f,j}+N_{b,j}} \left[
\frac{N_{f,j}N_{b,j}}{N_{f,j}+N_{b,j}}  \right]^{1/2}  \nonumber \\
\nonumber
M_{j} &=&
\frac{|c_{j}|}{(1+c_{j}^{2}+h(M_{\ell\ell},\theta,P_{T}))} \nonumber 
\end{eqnarray}
where we have used $\Delta N_{f,j}= (N_{f,j})^{1/2}$ and  $\Delta
N_{b,j}= (N_{b,j})^{1/2}$. 

The above expression shows that for case in which we have  same
number of
events in each of the $|\cos\theta_{j}|$  bins,  the error in the
extracted measurement of 
$A_{fb-j}^{total}$ from the data in  a specific  $|c_{j}|=
|\cos\theta_{j}|$ bin is equal to the error
of the measured  asymmetry in the bin divided by a factor  $M_{j} =
\frac{|c_{j}|}{(1+c_{j}^{2}+h(\theta))}$.
This factor comes from the fact that the extracted total
forward-backward asymmetry is more sensitive to
events at large $|c_{j}|= |\cos\theta_{j}|$.

We now convert the procedure to event weight technique.  We define
$A_{fb-j}^{total}=(3/8) (N_{A,j}/N_{B,j})$. 
\begin{eqnarray}
N_{f,j}&=& N_{A,j}(1+c_{j}^{2}+h(\theta)) + |c_j|N_{B,j}\\  \nonumber
N_{b,j}&=& N_{A,j}(1+c_{j}^{2}+h(\theta)) -|c_j|N_{B,j} \nonumber
\end{eqnarray}
From which we get: 
\begin{eqnarray}
N_{A,j} &=&  \frac {N_{f,j} }{2(1+c_{j}^{2}+h)} +    \frac {N_{b,j}}
{2(1+c_{j}^{2}+h)} \\ \nonumber
N_{B,j} &=&   \frac {N_{f,j}} {2|c_{j}|} -   \frac {N_{b,j}}
{2|c_{j}|}  \nonumber
\end{eqnarray}

In order to  properly weight events for different  $|\cos\theta_{j}|$
bins by the inverse of the square of the
error for each bin we  multiply the above expressions  by
\begin{equation}
M_{j} ^{2}= \frac{c_{j}^{2}}  {(1+c_{j}^{2}+h(M_{\ell\ell},\theta,
P_{T}) )^{2}}   \nonumber
\end{equation}
and get : 
\begin{eqnarray}
\label{ang1}
z_{1,j} &=&   \frac{1}{2}  \frac {c_{j}^{2}}
{(1+c_{j}^{2}+h(\theta))^{3}}  \\ \nonumber
z_{2,j} &=&   \frac{1}{2}  \frac {|c_{j}|}  {(1+c_{j}^{2}+h(\theta
))^{2}}  \\ \nonumber
A_{1} &=&    N_{f,j}  \cdot (z_{1,j} ) \\ \nonumber
A_{2} &=&    N_{b,j}  \cdot (z_{1,j}) \\ \nonumber
B_{1} &=&    N_{f,j}   \cdot (z_{2,j} )\\ \nonumber
B_{2} &=&    N_{b,j}   \cdot (z_{2,j} )\\ \nonumber
\left[ \Delta A_{1} \right]^{2}  &=&    N_{f,j}  \cdot z_{1,j} ^{2}\\
\nonumber
\left[ \Delta A_{2} \right]^{2}  &=&      N_{b,j}  \cdot z_{1,j}^{2}
\\ \nonumber
\left[ \Delta B_{1} \right]^{2}  &=&    N_{f,j}  \cdot z_{2,j}^{2} \\
\nonumber
\left[ \Delta B_{2} \right]^{2}  &=&    N_{b,j}  \cdot z_{2,j}^{ 2}\\
\nonumber
A_{j} &=&    N_{f,j} (z_{1,j} )+    N_{b,j} (z_{1,j}) \nonumber \\
&=&  A_{1} + A_{2}\\
B_{j} &=&    N_{f,j}  (z_{2,j} )-   N_{b,j} (z_{2,j} )\nonumber \\
&=& B_{1}  - B_{2}  \\
A_{fb-j}^{total} &=& \frac {3}{8}  \frac{B_{j} }{A_{j}}  =  \frac
{3}{8}  \frac{B_{1}  - B_{2}} {A_{1} + A_{2}}  \nonumber
\end{eqnarray}

Now  $\Delta A_{1}$ is $100\%$ correlated with $\Delta B_{1}$   and
$\Delta A_{2}$ is $100\%$ correlated with $\Delta B_{2}$ . 
We handle these correlations  as follows.  
\begin{eqnarray}
\label{ang2}
\Delta A_{1} & = &  \Delta B_{1} \cdot \frac{A1}{B1} \\ \nonumber
\Delta A_{2} & = &  \Delta B_{2} \cdot \frac{A2}{B2 } \\ \nonumber
 \left[ \Delta A_{fb-j} ^{total}\right] ^{2}  &=&   \left[  \frac
{3}{8} \right] ^{2}
\frac {1}  {(A_{1}+A_{2})^{4}}  \left[ E_{1}^{2} + E_{2}^{2}  \right]
\\  \nonumber
E_{1}^{2} & = &    \frac  {\left[ \Delta B_{1}
\right]^{2}}{B^{2}_{1}} (A2  B1 +A1  B2 )^{2}\\ \nonumber
E_{2}^{2}&=&   \frac {\left[ \Delta B_{2} \right]^{2}}{B^{2}_{2}}
(A2  B1 +A1  B2 )^{2} \nonumber
\end{eqnarray}

Table \ref{table2}  shows the results of a numerical example for an
asymmetry $A_{fb}$ = 0.6 measured with 1000 events (we assume h=0).
Here  $A_{fb}^{total}$ is measured in  ten bins of  $|\cos\theta|$.
 In each  $|\cos\theta_j|$ bin  we compare the error from the
standard error formula, and the error from the event weighting
formula. As expected, the two methods yield identical results.
 
 In addition, we show a comparison of  the average $A_{fb}^{total}$
for the ten $|\cos\theta_j|$ bins 
   calculated in three different ways. The error in the average
extracted from  a least square fit to the 10  $A_{fb-j}^{total}$
values (0.0196)  is close to the error in the  average  determined
from the weighted sum of all the events (0.0210).   The error in
$A_{fb}^{total}$  from the weighted sum of all the events  is $20\%$
lower than the error of 0.0253 obtained  from a simple count of all
forward and backward events.

For the case of low statistics, we can
use the event weighting technique to  combine all the events at all
value of  $|\cos\theta|$  (we do not need to bin the events in
$|\cos\theta|$).   The  following are the formulae  to extract
the best value and error from the entire range in $|\cos\theta|$
using the event weighting technique :

\begin{eqnarray}
z_{1,j} &=&   \frac{1}{2}  \frac {c_{j}^{2}}  {(1+c_{j}^{2}+h(\theta,
P_{T}) )^{3}}  \\  \nonumber
z_{2,j} &=&   \frac{1}{2}  \frac {|c_{j}|}  {(1+c_{j}^{2}+h(\theta,
P_{T}) )^{2}}  \\ \nonumber
    N_{total} & =&  \sum_{all-events}\left[  1 \right] \\ \nonumber
A_{1} &=&    \sum_{forward-events}  \left[  z_{1,j} \right]  \\
\nonumber
A_{2} &=&   \sum_{back-events}  \left[ z_{1,j} \right] \\ \nonumber
B_{1} &=&   \sum_{forward-events}   \left[   z_{2,j} \right] \\
\nonumber
B_{2} &=&   \sum_{back-events}   \left[  z_{2,j} \right] \\ \nonumber
\left[ \Delta A_{1} \right]^{2}  &=&   \sum_{forward-events}   \left[
z_{1,j} ^{2}\right] \\ \nonumber
\left[ \Delta A_{2} \right]^{2}  &=&    \sum_{back-events} \left[
z_{1,j}^{2}\right]  \\ \nonumber
\left[ \Delta B_{1} \right]^{2}  &=&   \sum_{forward-events} \left[
z_{2,j}^{2}\right]  \\ \nonumber
\left[ \Delta B_{2} \right]^{2}  &=&   \sum_{back-events}\left[
z_{2,j}^{ 2}\right] \\ \nonumber
A &=&  A_{1} + A_{2}\\ \nonumber
B &=&  B_{1}  - B_{2} \\ \nonumber
\left[ A_{fb}\right]^{total}&=&   \frac {3}{8} \frac{B} {A}  =  \frac
{3}{8}  \frac{B_{1}  - B_{2} } {A_{1} + A_{2}}\\ \nonumber
\Delta A_{1} & = &  \Delta B_{1} \cdot \frac{A1}{B1} \\  \nonumber
\Delta A_{2} & = &  \Delta B_{2} \cdot \frac{A2}{B2 } \\  \nonumber
 \left[ \Delta A_{fb} ^{total}\right] ^{2}  &=&   \left[  \frac
{3}{8} \right] ^{2}
\frac {1}  {(A_{1}+A_{2})^{4}}  \left[ E_{1}^{2} + E_{2}^{2}  \right]
\\  \nonumber
E_{1}^{2} & = &    \frac  {\left[ \Delta B_{1}
\right]^{2}}{B^{2}_{1}} (A2  B1 +A1  B2 )^{2}\\  \nonumber
E_{2}^{2}&=&   \frac {\left[ \Delta B_{2} \right]^{2}}{B^{2}_{2}}
(A2  B1 +A1  B2 )^{2}  \nonumber
\end{eqnarray}

Note that since we add up the forward and backwards events in
separate sums, the weighting
factors  $z_{1,j}$ and $z_{2,j}$ are functions of the absolute value
$|\cos\theta|$.

Table \ref{table3} shows a numerical example of measuring  $A_{fb}$
for
different acceptance ranges in  $|\cos\theta|$ (we assume h=0).
The example is for an asymmetry $A_{fb}$ = 0.6  and  an a sample of
1000 events
(for  $0<|\cos\theta|<1$). 
 For each range of acceptance in $|\cos\theta|$ we show the error in
$A_{fb}$ from a simple count,  the error from the event weighting
procedure, and  the error from fitting
bins in $|\cos\theta|$ (which is very similar to event weighting).
Also shown is the improvement factor  in the error when the event
weighting procedure 
is used (versus a simple count).  For a typical range of acceptance
in $|\cos\theta|$, using the simple weighting formula leads to  more
than $20\%$ reduction in the error. 

Note that when we use the angular distribution weights, the extracted
$A_{fb}^{total}$  is automatically corrected for the acceptance in
$|\cos\theta|$ (since the
acceptance cancels to first order). 
However, $N_{total}  =  \sum_{all-events}\left[  1 \right]$ is equal
to the observed
number of events and  is not corrected for acceptance. For
$N_{total}$,  an acceptance correction is needed
for the determination of  $d\sigma/dM_{\ell\ell}$    
  
   \begin{table*}[htb]
\caption { 
Proton-Proton Collisions: Numerical example of measuring  $A_{fb}$
integrating over various
ranges  of  $|\cos\theta|$ and various ranges of $|y|$ for the case
of a dilepton mass
of 450 $GeV/c^2$ at the LHC. Here we assume that the   
 asymmetry $A_{fb}$ = 0.6 is  measured with $10^{6}$ events for
values of    $|\cos\theta|$ 
and $0<|y|<2$.
Shown is the improvement factor in the  the error (over the simple
count method)
when we use event weighting in both  $|\cos\theta|$  and $w=MisID$ as
a function of $|y|$.  
For a typical range of acceptance in  $|\cos\theta|$ and $|y|$,  the
weighting formula leads to  a $40\%$ reduction in the error (over the
error obtained from a simple count).}
\label{table4}
\newcommand{\m}{\hphantom{$-$}}
\newcommand{\cc}[1]{\multicolumn{1}{c}{#1}}
\renewcommand{\tabcolsep}{0.75pc} 
   \begin{center}
\begin{tabular}{|c|cccccccccc|}
\hline
$y-range$  & 0-0.2 & 0-0.4 &   0-0.6 & 0-0.8 & 0-1.0 &0-1.2 &0-1.4 &
0-1.6 & 0-1.8 & 0-2.0 \\
\hline
$|\cos\theta|$  &  &  &   &  &  &  & & &  &  \\
$range$  &  &  &   &  &  &  & & &  &  \\
\hline\hline
0-0.1	&	1.00	&	1.12	&	1.14	&	1.15	&	1.16	&	1.17	&	1.18	&	1.21	&	1.23
&	1.26	\\
0-0.2	&	1.13	&	1.27	&	1.29	&	1.30	&	1.31	&	1.32	&	1.34	&	1.37	&	1.39
&	1.43	\\
0-0.3	&	1.17	&	1.31	&	1.34	&	1.35	&	1.36	&	1.37	&	1.38	&	1.42	&	1.44
&	1.48	\\
0-0.4	&	1.20	&	1.35	&	1.37	&	1.38	&	1.39	&	1.41	&	1.42	&	1.45	&	1.48
&	1.51	\\
0-0.5	&	1.23	&	1.37	&	1.40	&	1.41	&	1.42	&	1.43	&	1.45	&	1.48	&	1.51
&	1.54	\\
0-0.6	&	1.24	&	1.39	&	1.41	&	1.43	&	1.44	&	1.45	&	1.46	&	1.50	&	1.53
&	1.56	\\
0-0.7	&	1.25	&	1.40	&	1.42	&	1.43	&	1.45	&	1.46	&	1.47	&	1.51	&	1.53
&	1.57	\\
0-0.8	&	1.24	&	1.39	&	1.42	&	1.43	&	1.44	&	1.45	&	1.47	&	1.50	&	1.53
&	1.56	\\
0-0.9	&	1.23	&	1.37	&	1.40	&	1.41	&	1.42	&	1.44	&	1.45	&	1.49	&	1.51
&	1.55	\\
0-1.0	&	1.21	&	1.35	&	1.37	&	1.39	&	1.40	&	1.41	&	1.42	&	1.46	&	1.48
&	1.52	\\\hline
\end{tabular}
   \end{center}
\end{table*}

 \section{Combining $misID$ weighting and angular distribution
weighting in $pp$ collisions}
  
 In $pp$ collisions each event is can be characterized  by a $misID$
factor $w_{i}(|y_{i}|)$ which is related to the quark and antiquark
distribution ($x_{1,2} = (M_{\ell\ell}/\sqrt{s})e^{\pm |y|}$) 
   at its value of $|y_{i}|$. In addition, each event has a measured
value of  $|c_{j}|=|\cos\theta_{j}|$ . The expressions for combining
events with different  $|c_{j}|=|\cos\theta_{j}|$  and  $misID$
 $w_{i}$ values are given by:  
 
 \begin{eqnarray}
        k_{A,i} &=&    k_{1,i} -   k_{2,i} =(1-2w_{i})^{2}\\
\nonumber
k_{B,i} & =&    k_{1,i}  +   k_{2,i} =  (1-2w_{i}) \\  \nonumber
    N_{total} & =&  \sum_{all-events}\left[  1 \right] \\  \nonumber
z_{1,j} &=&   \frac{1}{2}  \frac {c_{j}^{2}}  {(1+c_{j}^{2}+h(\theta
))^{3}}  \\  \nonumber
z_{2,j} &=&   \frac{1}{2}  \frac {|c_{j}|}
{(1+c_{j}^{2}+h(\theta))^{2}}  \\  \nonumber
A_{1} &=&    \sum_{for-events}  \left[  z_{1,j}  k_{A,j}\right] \\
\nonumber
A_{2} &=&   \sum_{back-events}  \left[ z_{1,j}  k_{A,j}\right]   \\
\nonumber
B_{1} &=&   \sum_{for-events}   \left[   z_{2,j} k_{B,j} \right]  \\
\nonumber
B_{2} &=&   \sum_{back-events}   \left[  z_{2,j} k_{B,j}  \right]
\\  \nonumber
\left[ \Delta A_{1} \right]^{2}  &=&   \sum_{for-events}   \left[
z_{1,j} ^{2} k_{A,j}^2\right] \\  \nonumber
\left[ \Delta A_{2} \right]^{2}  &=&    \sum_{back-events} \left[
z_{1,j}^{2} k_{A,j}^{2}\right]  \\  \nonumber
\left[ \Delta B_{1} \right]^{2}  &=&   \sum_{for-events} \left[
z_{2,j}^{2} k_{B,j} ^{2}\right]   \\  \nonumber
\left[ \Delta B_{2} \right]^{2}  &=&   \sum_{back-events}\left[
z_{2,j}^{ 2} k_{B,j} ^{2}\right]      \\  \nonumber
A  &=&   A_{1} + A_{2}\\  \nonumber
B  &=&  B_{1}  - B_{2} \\  \nonumber
\left[ \Delta A_{1} \right] & = &  \left[ \Delta B_{1}\right] \cdot
\frac{A1}{B1} \\  \nonumber
\left[ \Delta A_{2} \right]& = &  \left[ \Delta B_{2}\right] \cdot
\frac{A2}{B2 } \\  \nonumber
A_{fb}^{total}&=&   \frac {3}{8} \frac{B} {A}  =  \frac {3}{8}
\frac{B_{1}  - B_{2} } {A_{1} + A_{2}}\\  \nonumber
 \left[ \Delta A_{fb} ^{total}\right] ^{2}  &=&   \left[  \frac
{3}{8} \right] ^{2}
\frac {1}  {(A_{1}+A_{2})^{4}}  \left[ E_{1}^{2} + E_{2}^{2}  \right]
\\  \nonumber
E_{1}^{2} & = &    \frac  {\left[ \Delta B_{1}
\right]^{2}}{B^{2}_{1}} (A2  B1 +A1  B2 )^{2}\\  \nonumber
E_{2}^{2}&=&   \frac {\left[ \Delta B_{2} \right]^{2}}{B^{2}_{2}}
(A2  B1 +A1  B2 )^{2}  \nonumber
\end{eqnarray} 

Table \ref{table4} shows a numerical example of  the improvement in
the errors that  that can be realized 
by using the information for both the $misID$ ($w_{i}$) and
$|\cos\theta_{i}|$ on an event by event basis,
versus  the measurement of  $A_{fb}$  using a simple count of all
events for all values
of  $|\cos\theta|$ and $|y|$ (within the experimental acceptance).
We show the case  
 for an asymmetry $A_{fb}$ = 0.6 measured with $10^{6}$  $pp$ events
(over all  values of    $|\cos\theta|$ and $0<|y|<2$ ) for the case
of a dilepton mass
of 450 $GeV/c^2$ at the LHC.  The two dimensional table
shows  the improvement factor in the  the errors (over the simple
count method)
when we use event weighting in both  $|\cos\theta|$  and $MisID$ as a
function of $|y|$.  
For a typical experimental acceptance in $|\cos\theta|$ and $|y|$,
the weighting formula leads to  a $40\%$ reduction in the error (over
the error obtained from a simple count).

The following are the advantages for using the event weighting. 

\begin{enumerate}

 \item For   $pp$ or $\bar{p}p$ collisions, if  event weighting is
 done  only in   $|\cos\theta|$,  the 
 error  in the asymmetry is typically reduced by a factor of 1.2.
This
the method that should be used to  extract the quark
couplings from  the average integrated asymmetry 
over all values of  $|y|$.

 \item  For $pp$ collisions,  if the  event weighting is
be done in both $|\cos\theta|$ and $|y|$,  the error  in the
asymmetry is  reduced by  a factor of 1.4.
For  $\bar{p}p$  collisions if the weighting is  done in both
$|\cos\theta|$ and $|y|$,
there is also a further reduction in the error.

  \item  For   $pp$ or $\bar{p}p$ collisions if  event weighting is
 done  only in    $|\cos\theta|$ 
   the method provides the acceptance corrected  asymmetry without
applying any acceptance corrections for missing coverage in
$|\cos\theta|$ . The  $|\cos\theta|$ acceptance fully cancels to
first order.
  For proton-antiproton collisions,  this is
  only true if the acceptance is the same for positive and negative
values  of  $|\cos\theta$.   If it is not, then the small difference
  in the acceptance for positive and negative  $|\cos\theta$ (e.g.
dead channels) needs to be accounted for
  in a MC acceptance correction.   Effects like a displaced beam
vertex need to be corrected for using a MC
  simulation (for the proton-antiproton case).  However, for
proton-proton collisions, effects of the  differences
  in the detector between forward and backward hemispheres   or a
displaced beam spot completely cancel. This is  because
 for proton-proton collisions  the forward-backwards are defined with
respect to the rapidity of the dilepton, which can be either positive
or negative in the detector.   A correction for the  acceptance in
$|y|$ still needs to applied
 in order to extract  the average integrated asymmetry 
over all values of  $|y|$. This
the method that should be used to  extract the quark
couplings from  the average integrated asymmetry 
over all values of  $|y|$.

  \item If the  event weighting is
be done in both $|\cos\theta|$ and $|y|$) then both 
the $|\cos\theta|$ acceptance and  the  $|y|$ acceptance fully
cancel.
This is the
method that should be used in a search for new $Z'$ bosons. 
It should also be used for the extraction of the asymmetry
in the production of top-antitop pairs, for which the acceptance
is a complicated function of both $|\cos\theta|$ and $|y|$.  

\item  Only small corrections  need to be made are detector
resolution and radiative smearing effects. Most of these  correction
are already included if  an empirical fit  $w_{i} = f
(y_{\ell\ell,measured},M_{\ell\ell,measured})$  is used. 

\item Only very small  corrections need to be made for QCD
modification of  the angular distribution. Most of these corrections
are already included if we use an empirical fit  for
$h(\theta,M_{\ell\ell},P_T)$, as given in Appendix A. 

\item The method uses event weights which are functions of either
$|y|$ or $|\cos\theta|$. This is equivalent
to weighting the data (in bins in $|y|$ or $|\cos\theta|$ ) by the
error in the  $expected$ number
of events.  Therefore, this is equivalent\cite{bodekNIM} to using the
maximum likelihood method (which yields
the smallest error). However, the implementation of this technique is
much easier and does not
depend on a detailed modeling of the detector.  
\item  If the  event weighting is
be done in both $|\cos\theta|$ and $|y|$) then both 
the $|\cos\theta|$ acceptance and  the  $|y|$ acceptance cancel.
Therefore, for this case 
  the weighted sums for A1, A2, B1, B2, $\Delta B_{1}$, and  $\Delta
B_{2}$  from different run conditions, or different 
final state leptons (muon,electrons) or different  experiments (e.g.
Dzero and CDF or CMS and ATLAS) can be directly added to provide a
combined result. This is important when there are only a few events
in each detector (e.g. 
 Drell-Yan events at high mass or top-antitop events). 

 \end {enumerate}

 \section{Correction factors and systematic uncertainties in the
weighting procedure}

  For $pp$ collisions, the acceptance for forward and backward events
is equal because of symmetry.  Therefore the functional dependence of
the acceptance in $|\cos\theta|$  fully cancels (if the
  acceptance for positive and negative muons is the same).  For
$\bar{p}p$ collisions one needs to correct for a possible small
difference in the detector performance between the proton and
antiproton directions. If the  event weighting is
be done in both $|\cos\theta|$ and $|y|$ then both 
the $|\cos\theta|$ acceptance and  the  $|y|$ acceptance cancel.

\subsection{ backgrounds and systematic errors}

 The main experimental  background is QCD dijet events.  The QCD jet
background  is measured by statistically separating  isolated  muons
(or electrons)  from muons (or electron like objects) in jets on the
basis of the transverse energy profile   distributions in the
calorimeter \cite{jyhan_thesis} (e.g.  isolation energy variables).
 
 In general  QCD processes  are mediated via spin 1 gluon exchange
and therefore have the same angular distribution as Drell Yan events.
If such is the case, the $fractional$ QCD background is the same
at all values of $\cos\theta$ and is the same for positive and
negative
 values of   $\cos\theta$.   Therefore, in the expressions for the
asymmetry, the
 QCD background cancels in the numerator. Corrections for this
background only increases the level of the denominator.
 Therefore, a single multiplicative factor equal to  $1/(1-f)$ (where
f is the fraction
 of QCD background events) can be used to correct  the extracted
$A_{fb}$  for QCD background.
The  same multiplicative correction factor for QCD background  can be
use for all  three
 methods (simple count,  fit to bins in  $|\cos\theta|$ and event
weighting technique). Possible deviations
 from this assumption can be included in the systematic error. 
 
Another  background originates from electroweak (EW)  processes
($WW$, $WZ$, $W$+jets, $\tau^{+}\tau^{-}$,
 and $t\bar{t}$). This background is  generally estimated from a
Monte Carlo simulation. If the event
 weighting technique is used to extract the forward-backward
asymmetry, then the event weighting
 technique can be used on Monte Carlo samples for the Drell-Yan
signal and the  electroweak background
 processes to determine the shift in $A_{fb}$ from EW backgrounds.
Note that the contribution from  $\tau^{+}\tau^{-}$ events is very
small (and at high mass  $A_{fb}$ for $\tau^{+}\tau^{-}$ is similar
to $A_{fb}$ for   $\mu\mu$ and $e^+e^-$ events).

  As is generally the case in particle physics experiments,  the
procedure needs to be tested  on Monte Carlo simulated data to
determine the size of any pulls from the previously listed systematic
effects.  
    We note that this process to determine biases and pulls  needs to
be done for any procedure that is used to extract the
forward-backward asymmetry for $q\bar{q}$ processes  from data.   Our
procedure is designed to minimize these biases, but they still need
to be determined from a full scale Monte Carlo simulation.
 
  In $pp$ collisions at the LHC, there are also uncertainties from
antiquark distribution functions that affect the 
$misID$ probabilities (which are determined using a Monte Carlo
simulations).  The  PDF uncertainties 
are smaller than the statistical errors for large dilepton mass.
Therefore, this method is best applied to searches
for new resonances or contact interactions at large mass.  At large
mass the  leading order  ${\bar q}q$ diagram dominates and the signal
for new physics  is  local change in the asymmetry  as a function of
dilepton  mass. 

However,  for precision measurement of  Standard Model parameters at
the LHC (e.g. near the region of the $Z$ boson) only weighting in
$|\cos\theta|$  should be used. 
This is because at lower dilepton mass, gluon-quark processes are a
significant fraction of the 
 cross section, and the dilution correction is much more sensitive to
PDFs and higher order QCD terms. 
 At low mass, it is better to compare the measured asymmetry (using
$|\cos\theta|$  weighting) to 
 a complete NLO or NNLO QCD prediction.

     \section{Summary}

     We have shown that a simple event weighting technique can be
used to reduce the statistical error on the
     extracted  $\bar{q}q$  forward-backward asymmetry from Drell-Yan
events in $\bar{p}p$ and $pp$ collisions. In addition to reducing the
statistical error, the event weighting technique is  much less
sensitive to
     details of the experimental  acceptance.
    
    The   asymmetry data for a  dilepton mass above 300 $GeV/c^2$
(for both $pp$ and $\bar{p}p$ collisions)   can be used to search for
new $Z'$ bosons.

 \section{Appendix A - Derivation of the perturbative formula for the
angular coefficients.}

We now show  that for  the $q\bar{q} \to \gamma^{*}/Z \to
{\ell^+\ell^-} $ annihilation
diagram  the perturbative calculation indicates that 
  the angular coefficients  originates from a  kinematic effect which
is a 
simple function of the angle of the proton and antiproton beams in
the Collins-Soper frame.  
That angle is  also a  simple function  of the transverse momentum  
$P_{T}$ and mass $M_{\ell\ell}$ of the dilepton pair.

We start  with the case
of the emission of a single hard gluon by one of the partons. 
In this case, the dilepton has a finite value of $P_{T}$ but we
cannot determine  if the
gluon was emitted by the parton in the proton, or by the parton in
the antiproton. 
All we know is that the dilepton pair has a finite transverse
momentum in the laboratory
system which is equal to the transverse momentum of the gluon.

If the gluon is emitted
by the parton in the antiproton, then  the parton level
angular distribution   (given below in $\cos\theta'$)
is with respect to the direction of the proton (which defines the 
quark-antiquark axis.  If the gluon is emitted
by the parton in the proton, then  the parton level
angular distribution (given below in $\cos\theta'$) is with respect
to the direction of the antiproton
(which defines the quark-antiquark axis). 
The Collins-Soper frame splits this difference and takes the average
direction
(which happens to be the same as the laboratory $z$ axis for
colliding beam experiments).

In the derivation below, 
all vectors are defined in the Collins-Soper frame.  
 We define unit vector $\bf k$ along the $z$ axis, 
 unit vector $\bf i$ along the $x$ axis of the Collins-Soper
 frame (which is defined to be along  the initial direction of the
transverse momentum of
the dilepton pair in the laboratory system). We also define a
perpendicular unit vector  $\bf j$ along the $y$ axis. 
Here $\theta$ is the angle with respect to the $z$ axis, and $\phi$
is defined
in the $x-y$ plane. 

 In the Collins-Soper frame, the
Drell-Yan pair is stationary and  proton
and antiproton move towards the dilepton pair with an angle $\xi$
with respect to the $z$ axis.
  In the Collins-Soper, we see an  antiproton moving from the
left at an angle $\xi$ with respect to the $z$  axis (in the
$x-z$ plane),   and  a proton
  moving from the right with
an angle$-\xi$ with respect to the  $z$ axis.  The  proton
and antiproton form an isosceles  triangle with
the two beam particles (which have equal momentum $p$)  intersecting
at an  angle $2\alpha= \pi-2\xi$.
Here  $~\cos~\alpha=~\sin~\xi$.  It can be easily shown that
\begin{eqnarray}
\sin^2~\xi  &=& \frac {P_T^2} {P_T^2 +M^2_{\ell\ell}}  \\  \nonumber 
\cos~\xi & =& \frac {M_{\ell\ell}} {(P_T^2 +M^2_{\ell\ell})^{1/2}}
\nonumber
\label{angles}
\end{eqnarray}
In the Collins-Soper frame, we  define a unit vector $\bf Z'$  along
the direction of one of the beam particles (e.g.  the proton).  We
define a unit vector $\bf r$ along the direction of Drell-Yan pair
decay positron. We then obtain the following relations.
\begin{eqnarray}
\label{cosine}
\bf{ Z'} & = &  ~\cos\xi~{ \bf{k}} + ~sin\xi~ \bf{i}\\ \nonumber 
{\bf {r}} &= & ~\sin\theta ~\cos\phi ~{\bf {i}} + ~sin\theta ~sin\phi
~{\bf {j}} +~\cos\theta ~{\bf {k}}\\ \nonumber
\cos\theta'&= & \bf{r} \cdot \bf{Z'}\\ \nonumber 
&= &   ~\cos\xi ~\cos\theta+\sin\xi ~\sin\theta ~\cos\phi
\end{eqnarray}
For the case for which the gluon was emitted by the parton in the
antiproton, the angular distribution of the decay positron with
respect to the $Z'$ proton axis
(which defines the quark-antiquark axis)  is given
by the simple expression. 

\begin{eqnarray}
\frac{d\sigma}{d \cos(\theta)} &=& A \left[  ( 1 + \cos^{2} \theta')+
\frac{B'}{A} \cos\theta' \right] 
\label{parton}
\end{eqnarray}
(if the gluon was emitted by the proton, then the above distribution
is with respect to the
direction of the antiproton which defines
the quark-antiquark axis).   Therefore, 
the distributions for the proton direction and for the antiproton
direction
should be added. Each should be weighted by the probability
that the quark emitted the  gluon $P_q(x_1)$,   or  the probability
that the antiquark emitted a gluon,  $P_{\bar q}(x_2)$.  Here
$P_q(x_1)$+$P_{\bar q}(x_2)$=1.
For proton-antiproton collisions, the two probabilities should be
equal.  For
proton-proton collisions, they could be different.

The proton direction is at an angle  $\xi$. The antiproton direction
 angle is  $-\xi$.  If we plug in  the expression for $ \cos \theta'$
from equation \ref{cosine} into
equation \ref{parton}, we get  
\begin{eqnarray}
\frac{d\sigma}{d \cos(\theta' )} &=&  term_q + term_{\bar q} \\
\nonumber
term_q &=& A~ P_q(x_1) [ 1 + \cos^{2} \theta ~\cos^2\xi \\
\nonumber 
 &+& ~\sin^2\theta ~\sin^2\xi ~\cos^2\phi   + \\ \nonumber
 &+& ~2 \cos\xi ~\cos\theta sin\xi ~\sin\theta ~\cos\phi \\ \nonumber
&+&~\frac{B'}{A} ~\cos\xi ~\cos\theta  \\ \nonumber
&+&~sin\xi ~\sin\theta ~\cos\phi] \\ \nonumber
term_{\bar q} &=& C P_{\bar q}(x_2) [ 1 + \cos^{2} \theta ~\cos^2\xi
\\   \nonumber 
 &+& ~\sin^2\theta ~\sin^2\xi ~\cos^2\phi   + \\ \nonumber
 &-& ~2 \cos\xi ~\cos\theta sin\xi ~\sin\theta ~\cos\phi \\ \nonumber
&+& \frac{B'}{A} ~\cos\xi ~\cos\theta \\ \nonumber
&-& \frac{B'}{A}~sin\xi ~sin\theta ~\cos\phi]
\end{eqnarray}
 Adding $term_q$ and  $term_{\bar q}$  and defining 
 $\Delta P = (P_q(x_1)-P_{\bar q}(x_2)) \approx 0 $ we get
 \begin{eqnarray}
 \label{temp}
 \frac{d\sigma}{d \cos(\theta)} &=& A ~[ 1 + \cos^{2} \theta
~\cos^2\xi \\   \nonumber 
 &+&\sin^2\theta ~\sin^2\xi ~\cos^2\phi   + \\ \nonumber
 &+& {\Delta P}~ 2 \cos\xi ~\cos\theta ~sin\xi ~\sin\theta ~\cos\phi
\\ \nonumber
&+&\frac{B'}{A} ~\cos\xi ~\cos\theta  \\ \nonumber
&+&~{\Delta P} \frac{B'}{A}~sin\xi ~sin\theta ~\cos\phi] 
\end{eqnarray}
using trigonometric identities 
 \begin{eqnarray}
 \sin2\theta &=& 2 ~sin\theta ~\cos\theta \\ \nonumber
\cos^2\phi &=& \frac{1}{2} (1+\\cos2\phi) \\ \nonumber
\end{eqnarray}
we rewrite equation \ref{temp} as:
 \begin{eqnarray}
 \label{final}
 \frac{d\sigma}{d \cos(\theta)} &=& A [ 1 + \cos^{2} \theta)   \\
\nonumber 
  &+& ~\frac {\sin^2\xi}{2} (1-3 \cos^2\theta ) \\ \nonumber
    &+&~{\Delta P}~ \cos\xi \sin\xi ~\sin{2\theta} \cos\phi \\
\nonumber
 &+& ~\frac {\sin^2\xi}{2}~\sin^2\theta\cos^2\phi  \\ \nonumber
 &+&~{\Delta P} \frac{B'}{A}~\sin\xi ~\sin\theta \cos\phi \\ \nonumber
&+& ~\frac{B'}{A}\cos\xi \cos\theta  \nonumber ]
\end{eqnarray}
The general expression
for the total differential cross section for dilepton pair production
(e.g. Drell-Yan,  $Z's$ or $W's$)  
in proton-antiproton (or proton-proton) collisions is given by:
\begin{eqnarray}
{\sigma} &=& A~[(1+\cos^2{\theta})  \nonumber \\
&+& \frac{1}{2}A_0(1-3\cos^2{\theta})
+ A_1\sin{2\theta}\cos{\phi} \nonumber \\
&+&\frac{1}{2}A_2 \sin^2{\theta}\cos{2\phi}
+ A_3\sin{\theta} \cos{\phi} \nonumber \\
&+& A_4 \cos{\theta} +A_5  \sin{2\theta}\sin2\phi  \nonumber \\
&+& A_6 \sin{2\theta}\sin{\phi}
+ A_7 \sin{\theta}\sin{\phi}]
\label{AngleFuncA}
\end{eqnarray}
where $A_4= \frac {B}{A}$. Comparing equation
\ref{final} to equation \ref{AngleFuncA} we get:
\begin{eqnarray}
A_0  &= & ~ \sin^2\xi = \frac {P_T^2} {P_T^2 +M^2_{\ell\ell}}  \\
\nonumber 
 A_1  &=&~{ \Delta P}~  \cos\xi ~ \sin\xi = {\Delta P}~ \frac {P_T
M_{\ell\ell}} {P_T^2 +M^2_{\ell\ell}} \approx 0 \\ \nonumber
 A_2&= & ~ \sin^2\xi = \frac {P_T^2} {P_T^2 +M^2_{\ell\ell}}  \\
\nonumber 
A_3 &=&~{\Delta P}  \frac{B'}{A}~\sin\xi  =  {\Delta P}  \frac{B'}{A}
\frac {P_T} {(P_T^2 +M^2_{\ell\ell}) ^{1/2}}  \approx 0  \\
\nonumber 
A_4 &=& \frac {B}{A} =\frac{B'}{A} ~\cos\xi  =   \frac{B'}{A} \frac
{M_{\ell\ell}} {(P_T^2 +M^2_{ \ell\ell})^{1/2}} \\ \nonumber
\frac {B}{B'} & = &   \frac {M_{\ell\ell}} {(P_T^2
+M^2_{\ell\ell})^{1/2}} \\ \nonumber
A_5 &=& A_6=A_7=0  \nonumber
\end{eqnarray}
Where $B$ is the $\cos\theta$ angular coefficient for quarks bound in
the proton, and $B'$ is
the $\cos\theta$ angular coefficient for free quarks.
As far as we know, this is the first closed form derivation of the
$QCD$ corrections
to the forward backward asymmetry (i.e. the expression above for
$A_4$ and $\frac {B}{B'}$). 
The above perturbative  formulae for $A_0$ and $A_2$ for the
annihilation diagram
for Drell-Yan production  
 was first derived  by  Collins  in 1979\cite{azeroLO}. The 
 perturbative  expressions for $A_0$, $A_1$  and $A_2$ and  $\Delta
P$ (but not  for  $A_3$ and $A_4$) were 
 also derived in a paper by Daniel Boer and  Werner Vogelsang in
2006)\cite{azeroLO}.   The above expressions for  $A_0$ and $A_2$
were shown to hold for the quark-antiquark annihilation diagram for
all orders (re-summation) by Berger et. al. in 2007\cite{berger}.  

In conclusion, if the $Z/DY$ cross section is dominated by
quark-antiquark annihilation (as expected)
the perturbative calculation gives predictions for  $A_0$ ($=A_2$)
and $A_4$ as the only surviving non zero angular coefficients. 
The simple QCD predictions for  the angular coefficients  are:  
\begin{eqnarray}
h(\theta,M_{\ell\ell},P_T) & = &   \frac{1}{2}A_0(1-3\cos^2{\theta})
\\ \nonumber
A_0  &= & ~\sin^2\xi \\ \nonumber
\sin^2\xi & = &  \frac {P_T^2} {P_T^2 +M^2_{\ell\ell}}\\   \nonumber 
A_4&=& \frac{B}{A} =  \frac{B'}{A}~\cos\xi \\ \nonumber
\frac {B}{B'} & = & ~\cos\xi \\ \nonumber
\cos~\xi & =& \frac {M_{\ell\ell}} {(P_T^2 +M^2_{\ell\ell})^{1/2}}
\nonumber
\end{eqnarray}

The angular distributions predicted  by standard Monte Carlo
generators
such as  {\small PHYTIA}\cite{pythia},  ResBos
re-summation\cite{ResBos}  and
Vector Boson Production and Decay in Hadron Collisions Resummation
VBP\cite{Vbp}
for  $Z$ boson production at the Tevatron are well
 described by the above formulae.

 On average,  both QCD correction terms are very small. 
  For Drell-Yan events  the average ratio $P_T/M_{\ell\ell}$ is about
0.13 (e.g. 12 GeV $P_T$
 at the $Z$ mass of 90 GeV). This implies that on average
$<A_0(P_T)>= 0.018$, and   $<\frac {B}{B'}(P_T>=0.991$.

\subsubsection{Including the effect of photon radiation by final
state leptons}

Here, we derive an approximate formula for including the effect of
final state photons.
The more exact way  is to run the Monte Carlo, with and without photon
radiation and find a correction to the final result.  
empirical correction
radiation.

We assume that final state radiation smears the measured angle of the
final state electrons by an $RMS$  angle $\pm \delta$ in the
$\theta$ direction,  and the same
angle $\pm \delta$  in $\phi$.  The $\theta$ smearing is equivalent
of taking the average of the above expressions
with  $\xi + \delta$ and  $\xi - \delta$.    The $\phi$ smearing is
equivalent of taking the average of the above expressions
with  $2\phi + \delta$ and  $2\phi - \delta$. Therefore, the effect
of final state photon
radiation can be approximated by:

\begin{eqnarray}
A_0^{rad}  &= & \frac{1}{2}[ \sin^2(\xi +\delta) +  \sin^2(\xi
-\delta) ]  \\  \nonumber 
     &= &  \sin^2(\xi)~\cos^2 \delta  +  \cos^2 \xi !~\sin^2 \delta
\\  \nonumber 
     A_2 ^{rad}&=& [ \sin^2 \xi~ \cos^2 \delta +  \cos^2 \xi~ \sin^2
\delta] \\ \nonumber
    & & \times  [\frac{1}{2}[\cos(2\phi + \delta)+ \cos(2\phi +
\delta) ]\\ \nonumber 
     & = &   [ \sin^2 \xi ~\cos^2 \delta +  \cos^2 \xi~ \sin^2 \delta
] ~\cos\delta \\ \nonumber
     & = & A_0^{rad}\cos\delta  \\ \nonumber
 A_4^{rad}  &=& \frac{B'}{A} ~\frac{1}{2}[  \cos(\xi +\delta)
\cos(\xi -\delta) ] \\ \nonumber
 &=& frac{B'}{A} \cos \xi !~\cos \delta   \nonumber
 \label{radformula}
\end{eqnarray}

The above formula approximates  the effects of photon radiation using
one effective parameter $\delta$. That
parameter depends on the specific experimental cuts and channels.
For example,
if the final state is composed of two electrons, some of the photons
are merged in with
the electron shower in the calorimeter and help restore part of  the
original
electron energy and direction.   If the final state is a dimuon, the
radiated
photons are not included in the muon energy.  Therefore $\delta$  is
only
an effective parameter that needs to be determined from fits to a
Monte Carlo simulation. 
Probably, the simplest way is to include the corrections for the
effect of photon radiation
as part of an overall MC radiative correction.  Alternatively, we
could extract $\delta$ from a fit to the MC and
parametrize the  QCD+QED  corrections to the angular distribution by:

\begin{eqnarray}
h(\theta,M_{\ell\ell},P_T) & = &
\frac{1}{2}A_0^{rad}(1-3\cos^2{\theta})  \\ \nonumber
 A_0^{rad}  &= &  \sin^2 \xi ~ \cos^2 \delta +  \cos^2\xi ~\sin^2
\delta  \\  \nonumber 
 A_4^{rad}  &=& \frac {B^{rad}}{A} =~\frac{B'}{A} \cos \xi ~\cos
\delta  \\  \nonumber
 \frac {B^{rad}}{B'} &=&  ~\cos \xi ~ \cos \delta \\ \nonumber
\sin^2\xi & =&  \frac {P_T^2} {P_T^2 +M^2_{\ell\ell}}\\   \nonumber 
\cos~\xi & =& \frac {M_{\ell\ell}} {(P_T^2 +M^2_{\ell\ell})^{1/2}}
\nonumber
\end{eqnarray}

For $\delta$ of order 0.1, the effect of QED photons is of the same
magnitude
as the effect of the QCD corrections.


\begin{thebibliography}{9}
\bibitem{cdf_coord}
We define  ($\theta$,$\phi$,$z$), where $\theta$ is the polar angle
relative to
the quark direction  (the $+z$ axis), and $\phi$ the azimuth.
For a lepton ($e^{+}+e^{-}$ or  $\mu^{+}+\mu^{-}$)   pair
 with transverse momentum $P_{T}=P\sin\theta$,  energy
$E_{T}=E\sin\theta$, 
 the rapidity  $y=\frac{1}{2}ln\frac{E+P_{z}}{E-P_{z}}$. Here, 
$P$ and $P_{z}$ are the magnitude and z component of the momentum,
and E is the energy of the lepton pair.
\end{thebibliography}
\end{document}